%% file: main.tex
\documentclass[conference]{IEEEtran}
\usepackage{bbm}
\usepackage{epsfig}
\usepackage{amssymb} 
\usepackage[tbtags]{amsmath} 
\usepackage{graphics,eepic,epic}
\usepackage{latexsym}
\usepackage{euscript}
\usepackage{subfigure}
\usepackage{graphics,eepic,epic,psfrag}

\usepackage[numbers]{natbib}
\usepackage{notoccite}

\usepackage[english]{babel}
\usepackage{fancyhdr}
\usepackage{lastpage}
\input{styles/gww_defs}


\input{styles/defs}

\usepackage{float}

\begin{document}
% paper title
\title{A Hierarchical Graph Signal Processing Approach to Inference from Spatiotemporal Signals}%, a seizure detection case study}

% author names and affiliations
% use a multiple column layout for up to three different
% affiliations

\author{\IEEEauthorblockN{Nafiseh Ghoroghchian, Stark C. Draper, and Roman Genov }\\
	\IEEEauthorblockA{Department of Electrical and Computer  Engineering, University of Toronto, Toronto, ON, Canada}
	\IEEEauthorblockA{ nafiseh.ghoroghchian@mail.utoronto.ca, david.m.groppe@gmail.com, stark.draper@utoronto.ca, and roman@eecg.utoronto.ca }}

\maketitle
\begin{abstract}
Motivated by the emerging area of graph signal processing (GSP), we introduce a novel method to draw inference from spatiotemporal signals. Data acquisition in different locations over time is common in sensor networks, for diverse applications ranging from object tracking in wireless networks to medical uses such as electroencephalography (EEG) signal processing.
In this paper we leverage novel techniques of GSP to develop a hierarchical feature extraction approach by mapping the data onto a series of spatiotemporal graphs. Such a model maps signals onto vertices of a graph and the time-space dependencies among signals are modeled by the edge weights.
Signal components acquired from different locations and time often have complicated functional dependencies. Accordingly, their corresponding graph weights are learned from data and used in two ways. First, they are used as a part of the embedding related to the topology of graph, such as density. Second, they provide the connectivities of the base graph for extracting higher level GSP-based features. The latter include the energies of the signal's graph Fourier transform in different frequency bands. 
We test our approach on the intracranial EEG (iEEG) data set of the Kaggle epileptic seizure detection contest. In comparison to the winning code, the results show a slight net improvement and up to 6 percent improvement in per subject analysis, while the number of features are decreased by 75 percent on average.

\end{abstract}

\section{Introduction}
A spatiotemporal signal is acquired from different locations and times. The often high dimensionality of the signal gives rise to the need to extract lower dimensional features. 
Graph signal processing (GSP) \cite{main_GSP} exploits graphical structure in data to realize more efficient and effective signal prossesing and inference. GSP adapts conventional signal processing notions to the graph domain, including graph frequency analysis \cite{main_GSP}, sampling, and filter design \cite{new_survey}.

Epilepsy is one of the most common neurological disorders in the world. Typical pharmaceutical or surgical treatments are not effective for all patients. As an alternative, medical devices have been developed to detect seizures and apply a therapeutic procedure such as deep brain stimulation \cite{kassiri2017closed}. Currently available devices suffer from high false alarm rate due to their simplistic seizure detection techniques. This paper is motivated by the need for more effective and efficient solutions. %\cite{}
The objective of epileptic seizure detection is to distinguish ictal period (during which seizures occur) from interictal (in which no seizures occur) based on EEG data. 
%Fig.~\ref{iEEG_data}. 
Despite a relatively long history of research on this topic, achieving near-perfect sensitivity and specificity for all patients in an efficient manner remains a challenge.
%\begin{figure}
%\centering 
%\includegraphics[width=8cm]{iEEG_sample.jpg}
%\caption{ iEEG sample data \cite{kaggle_data}}
%\label{iEEG_data}
%\end{figure}
Temporal and frequency-based features have been widely applied in epileptic seizure prediction and detection \cite{spectral_power_features,hill_winning_kaggle}. During seizures, there is synchronization across parts of the brain \cite{spatial_epilepsy}. Hence, we can hope to extract useful spatial properties from EEG data, properties to which little attention has been paid. % ,time_freq_features
The most typical approach to spatial feature extraction is to concatenate frequncy-based and temporal features of all channels and let the classifier detect the spatial correlations \cite{concat_epilepsy}. 
Phase locking value (PLV) is one of the few attempts to draw out space-related features from a different and more compact perspective \cite{PLV,spatial_epilepsy}. In this paper, we use GSP to develop combined temporal, spectral and spatial features.

Incorporating GSP to spatiotemporal data has recently been considered. For example, the approach in \cite{dynamic_graphs} takes advantage of spatiotemporal dynamics of signal by using time-varying graph topologies. The identification of a cognitive control event based on EEG data, is used as an example in \cite{DGFT}, where the notion of windowed dynamic graph Fourier transform is introduced.
In \cite{SVARM_brain_connectivity} the authors propose an algorithm for graph learning of spatiotemporal signals. They used iEEG data for evaluating a few network topology metrics before and during seizures. However, their approach has a high computational load and includes data acquired from a single patient which makes validating the generality of their approach difficult. Their work is also limited to evaluation and comparison of the calculated graph topology metrics. They do not use classification to make inference on a large data set. 

The contributions of this paper are twofold.
First, we introduce a hierarchical architecture that produces topology and GSP-based features that capture space, time and frequency-based characteristics of spatiotemporal data to enhance the efficacy of decision making. 
Second, this is the first time that a GSP-based approach on a spatiotemporal graph structure has been applied to epileptic seizure detection using large-scale iEEG data. Unlike existing papers \cite{SVARM_brain_connectivity,DGFT}, we do not limit testing the proposed approach to a few examples. Rather, we prove the efficacy of our proposed method by testing over thousands of data samples from a set of eleven test subjects.%changed

The rest of the paper is organized as follows. Section~\ref{sec.graph} describes an approach to graph construction and edge weight learning. In Section~\ref{sec.system_model}, the hierarchical feature extraction architecture is described. In Section~\ref{sec.Feature_extraction}, we describe two feature sets. The first quantifies the notion of topological graph features, and the second is based on concepts from GSP. Section \ref{sec.numerical} presents the numerical results. Section \ref{sec.conclusion} concludes the paper.% wherein we use the proposed method to extract the features of iEEG data for the purpose of epileptic seizure detection
\section{Graph Construction and Learning}\label{sec.graph}
We consider weighted undirected graphs of the form $\mathcal{G}(\mathcal{V},\mathcal{E},W)$ where $\mathcal{V}$ is a set of vertices of cardinality $|\mathcal{V}|$, $\mathcal{E}$ is a set of edges and $W \in \mathbb{R}^{|\mathcal{V}|\times |\mathcal{V}|}$ is a weighted adjacency matrix. The weight of the edge between vertices $u$ and $v$ is denoted by $W_{u,v}$ and is typically derived from the correlation (similarity) between the vertices. There is no edge between vertices $u$ and $v$ if $W_{u,v} = 0$. We assume undirected graphs. So, $W$ is a symmetric matrix. $\mathcal{N}_u = \left\lbrace u,v| W_{u,v} \neq 0 \right\rbrace$ is the neighbourhood of the $u$th vertex.
A signal on a graph using (``flattened'') vector notation $\mathbf{x}$, is a mapping $\mathbf{x}:\mathcal{V} \rightarrow \mathbb{R}^{|\mathcal{V}|\times 1}$. 

The main focus of the paper is on spatiotemporal signals. In this paper we consider discrete signals where both time and space are discrete values. 
We construct a spatiotemporal graph by mapping both time and spatial locations to graph vertices. Throughout the paper we use the previous flattened notation (using a single number $u$ for indexing an element in a signal or a vertex in the graph, i.e. $x_u$) and the following ``spatiotemporal-separated'' notation interchangeably. $\mathbf{X} \in \mathbb{R}^{S\times T}$ is a spatiotemporal signal:
\begin{align}\label{X_define}
\mathbf{X} = \begin{bmatrix}
x_{1,1} & x_{1,2} & \cdots & x_{1,T} \\
\cdots & \cdots & \cdots & \cdots \\
x_{S,1} & x_{S,2} & \cdots & x_{S,T} \\
\end{bmatrix} 
\end{align}
where $S$ is the number of locations or channels (e.g. the number of electrodes of the iEEG), $T$ is the number of samples, and $|\mathcal{V}| = ST$.
The rows of $\mathbf{X}$ correspond to the time-series recorded on each channel. Each row can be defined on a ``temporal graph''. Additionally, each column of $\mathbf{X}$ is the signal in all locations in one time-slot which can be defined on a ``spatial graph''.
Hence, for a spatiotemporal graph (from which a spatiotemporal signal is mapped), there is a one-to-one function $\Phi: \mathcal{V}\rightarrow [S] \times [T]$ ($[S]\triangleq\left\lbrace 1,\cdots, S \right\rbrace$) to map each vertex to a pair of space-time indices: %changed 
$\Phi(u)=(i,k): u \in \mathcal{V}$.
It should be noted that throughout the paper, $u$ and $v$ correspond to indices of the vertices in the flattened version, $i$ (and $j$) are used as spatial, and $k$ is used as temporal indices corresponding to space-time-separated version.
The adjacency matrix of the whole spatiotemporal graph is defined as \cite{dynamic_graphs,SVARM_brain_connectivity}:
\begin{align}\label{total_W}
\!\!\!W = \begin{bmatrix}
A_1 &\!\!\!\! B_{1,1}^\prime &\!\!\!\! B_{1,2}^\prime &\!\!\!\! \cdots &\!\!\!\! B_{1,L}^\prime &\!\!\!\! 0 &\!\!\!\!\!\!\!\! \cdots &\!\!\!\! 0 \\
B_{1,1} &\!\!\!\! A_2 &	B_{2,1}^\prime	&\!\!\!\!\! \cdots &\!\!\!\! B_{2,L-1}^\prime &\!\!\!\! B_{2,L}^\prime &\!\!\!\!\!\!\! \cdots &\!\!\!\! 0\\
\cdots &\!\!\!\! \cdots &\!\!\!\!\cdots &\!\!\!\!\! \cdots &\!\!\!\!\cdots &\!\!\!\! \cdots &\!\!\!\!\!\!\!\cdots &\!\!\!\! 0\\
0 &\!\!\!\! \cdots& \!\!\!\!\cdots & \!\!\!\!0 &\!\!\!\!\!\!\!\! B_{T-L,L}& \!\!\!\!\cdots &\!\!\!\! B_{T-1,1} & \!\!\!\! A_T \\
\end{bmatrix}
\end{align}
where $A_k$ is the weights among location vertices (spatial graph) in time $k$, $B_{k,{l}}(i,j)$ contains the weights between spatial vertex $i$ in time $k-l$ and spatial vertex $j$ in time $k$ (temporal graph), and $B^\prime$ is the transpose of $B$. The temporal connectivities of the graph is limited to $L$ hops.

After setting the structure of graph, we need to calculate the adjacency matrix. 
We use a simple, yet effective graph learning strategy. We first center the spatial graph (removing the D.C. term):
\begin{align}
\mathbf{\overline{ X}}_{\text{sp}} \triangleq \mathbf{X} -
\mathbbm{1}_T
\begin{bmatrix}
\overline{{x}}_1^{(s)} & \overline{{x}}_2^{(s)} & \cdots & \overline{{x}}_T^{(s)}
\end{bmatrix}
\end{align}
where  $\mathbbm{1}_T$ is a vector of size $T\times 1$ and $\overline{{x}}_k^{(s)} \triangleq {\displaystyle\sum_{i\in[S]} {x}_{i,k}}/S$. We now define the $A_k$ to be used in \eqref{total_W} as:
\begin{align}\label{A_def}
A_k (i,j) \triangleq |{\overline{ x}}_{\text{sp}}(i,k){\overline{ x}}_{\text{sp}}(j,k)|.
\end{align}
We similarly center the temporal graphs, defining:
\begin{align}
\mathbf{\overline{ X}}_{\text{tl}} \triangleq \mathbf{X} - 
\begin{bmatrix}
\overline{{x}}_1^{(t)} &\overline{{x}}_2^{(t)} &\cdots&  \overline{{x}}_S^{(t)} 
\end{bmatrix}^\prime
\mathbbm{1}^\prime_S
\end{align}
where $\overline{{x}}_i^{(t)} \!\!\triangleq \!\!\!\!{\displaystyle\sum_{k\in[T]} \!\!{x}_{i,k}}/{T}\!$ and define the $\!B_{k,l}\!$ to be used in \eqref{total_W} as:
\begin{align}\label{B_def}
B_{k,l} (i,j) \triangleq |{\overline{ x}}_{\text{tl}}(i,k - l){\overline{ x}}_{\text{tl}}(j,k)|.
\end{align}
We take absolute value to make edge weights non-negative since: First, it matches better to graph notions \cite{journal_graph_laplacian_learning}. Second, our simulations (Section \ref{sec.numerical}) show that such definition provides better results. 
This graph learning process only contains simple summations and multiplications. Its order of computation is $\mathcal{O}(L S^2 T)$ which is the minimum possible order of computation of any graph learning algorithm (the same as the number of unknown values in adjacency matrix).
% The reason is that the same order of computation is required to at least announce the unknown values in adjacency matrix, i.e. $\displaystyle\sum_{k=1}^T \left[ \mathrm{size}( A_k)+\displaystyle\sum_{l=1}^L \mathrm{size}( B_{k,l})\right] = L S^2 T$. 
For later use we store the non-zero elements of the adjacency matrix in a tensor $\tau \in \mathbb{R}^{S\times S\times T\times (L+1)}$, where:
\begin{align}\label{tensor}
\tau_{i,j,k,l} = \left\{ 
  \begin{array}{l l}
   A_k (i,j)   &  \quad \mathrm{if} \quad  l = 0  \\
   B_{k,l} (i,j)  &  \quad   \mathrm{otherwise}.   \\    
  \end{array} \right.
\end{align}

There are multiple reasons behind suggesting such graph learning approach in comparison to other existing methods.
The most typical approach in GSP literature to determine graph connnectivity weights is thresholded Gaussian kernel weighting function which uses the distance between pairs of vertices \cite{main_GSP}. The distance can be Euclidean distance between the vertices which may not be a good measure in many applications. For instance, complex functional and axonal relations between different brain regions during motor movements result in high correlative activities among distant regions whereas the neighbouring regions can be uncorrelated \cite{functional_connectivity}. Accordingly, the need to learn the edge weights from data arises. 
There are many techniques for graph learning in the literature using various assumptions including ``diffusion'' \cite{diffusion_graph_learning}, ``global smoothness'' \cite{journal_graph_laplacian_learning} and ``time-varying observations'' \cite{SVARM_brain_connectivity}. However, they either do not match our intended application or suffer from high computational load.   
%There are also application-based approaches. For instance, the dynamic adjacency matrix of the spatial graph has been defined as the phase locking value of each pair of channels of EEG data to identify an event in a cognitive control experiment \cite{DGFT}. 

\section{Hierarchical feature extraction system}\label{sec.system_model}
\begin{figure}
\centering 
\includegraphics[width=8cm]{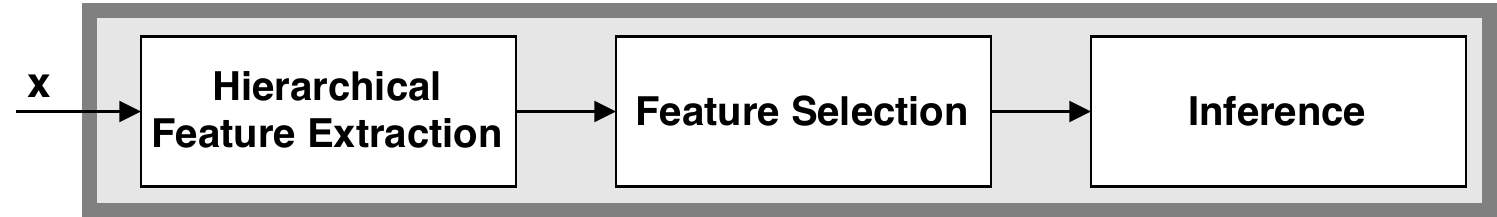}
\caption{Block diagram of the system }
\label{system_BD} 
\end{figure}

\begin{figure}
\centering 
\includegraphics[width=8cm]{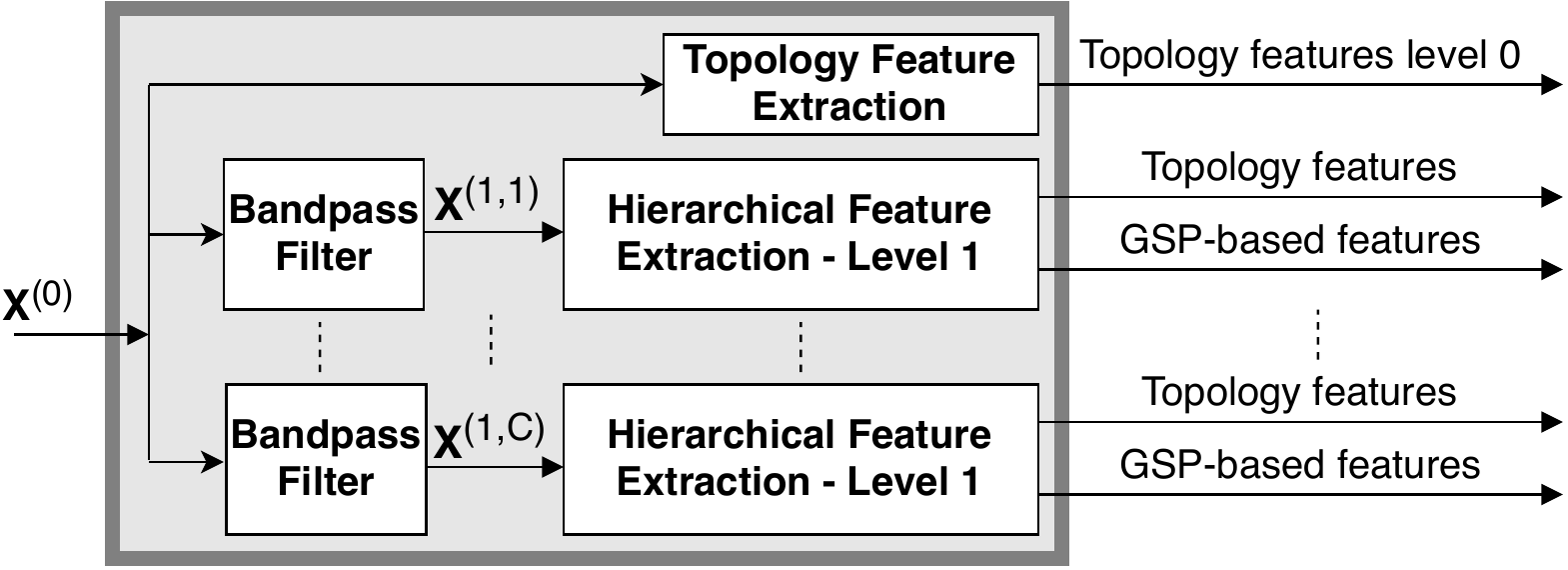}
\caption{Block diagram of the hierarchical feature extraction, level 0. }
\label{fig.FEL1} 
\end{figure}

\begin{figure}
\centering 
\includegraphics[width=8cm]{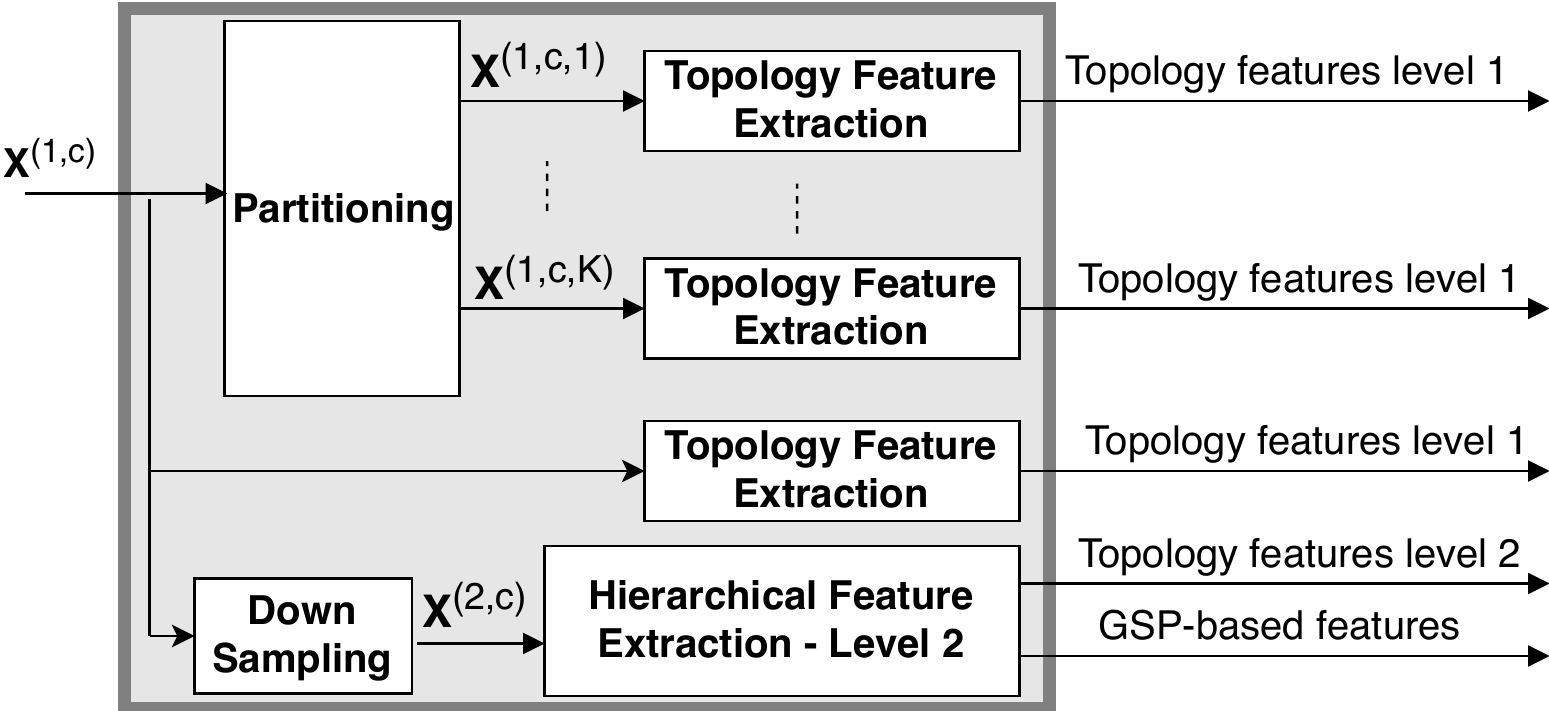}
\caption{Block diagram of the hierarchical feature extraction, level 1. }
\label{fig.FEL2} 
\end{figure}

\begin{figure}
\centering 
\includegraphics[width=6cm]{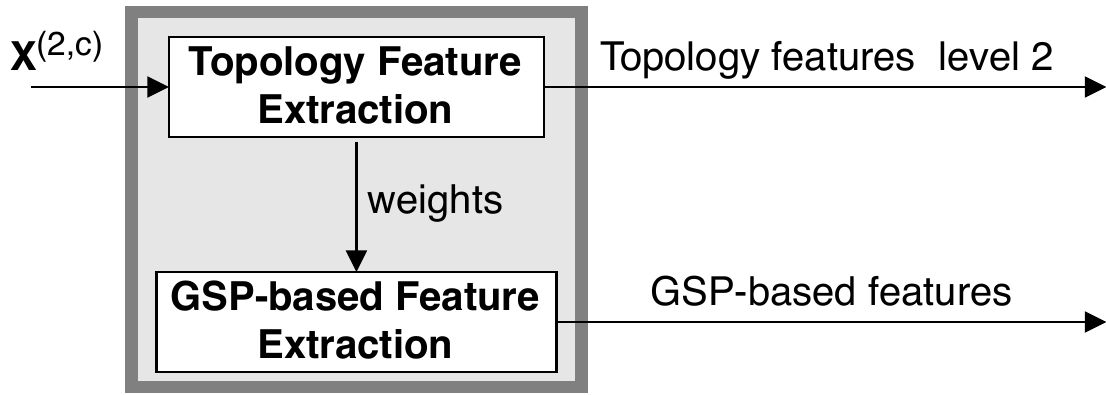}
\caption{Block diagram of the hierarchical feature extraction, level 2. }

\label{fig.FEL3} 
\end{figure}

The block diagram of the system is depicted in Fig.~\ref{system_BD}. We propose a 3-levels architecture for feature extraction  (figures \ref{fig.FEL1}-\ref{fig.FEL3}). Filtering, partitioning and down sampling are the main aspects that distinguish different levels.
Before elaborating on the details of hierarchical feature extraction system, we first explain two reasons behind suggesting such structure:

\noindent 
\textbf{1. High dimensionality:}
Constructing the whole two-dimensional ($2D$) adjacency matrix \eqref{total_W} from raw data can be a challenge due to its high dimensions. For instance, for $T=5000,S=55$, $563.5GB$ memory is needed only to store the matrix. As a result, we use the $4D$ tensor \eqref{tensor} to store only the non-zero indices of adjacency matrix. When $T$ is large, only a small $L$ results in a suitable degree of sparsity to justify using the tensor, rather than the $2D$ adjacency matrix. 

The $2D$ adjacency matrix is required for the two topology and GSP-based feature extraction blocks. However, we cannot work with the adjacency matrix drawn from the raw data because of its high dimensionality. Hence, for topology extraction, as will be described in Section \ref{sec.topology}, we map the adjacency matrix to lower dimensions. In addition, for GSP-based feature extraction (Section \ref{sec.GSP_features}), we work with the down sampled data.

\noindent 
\textbf{2. Capturing both coarse and fine features:}
3-stage down sampling and data partitioning, enable us to obtain fine (in levels 0 and 1) as well as coarse (in level 2) features. 

\noindent 
\textbf{Notation}: We use different notations for data in different levels. Let $\mathbf{X}^{(0)} \in \mathbb{R}^{S\times T_{0}}$ be the raw ``spatiotemporal-separated'' data (cf. \eqref{X_define}) in level $0$, while $\mathbf{X}^{(i,c)} \in \mathbb{R}^{S\times T_{i}}$ refers to the the down sampled and band-passed data where $i\in \lbrace 1,2\rbrace$ is the index of level and $c \in \left\lbrace 1,2,\cdots, {C}\right\rbrace $ is the band index.
%We first elaborate on the course of data through levels. Next, in Section \ref{sec.Feature_extraction}, the two main feature extraction building blocks, topology and GSP-based feature extractors, are described.

\noindent 
\textbf{Level 0}: The block diagram of this level is shown in Fig.~\ref{fig.FEL1}. The raw data $\mathbf{X}^{(0)} \in \mathbb{R}^{S\times T_{0}}$ is fed to a topology feature extraction block. The data is also fed to a filter bank of $C$ bandpass filters: $ \mathbf{X}^{(1,c)} = \text{BPF}_c(\mathbf{X}^{(0)})$.
Each idealized band-pass filter $\text{BPF}_c: \mathbb{R}^{S\times T_{0}}\rightarrow \mathbb{R}^{S\times T_{0}}$ is implemented by zeroing out the discrete-time Fourier series' components of the signal outside the frequency interval $[\omega_{c-1},\omega_{c})$ (BPF exploits \textit{conventional} Fourier transform. In level 2, the \textit{graph} Fourier transform is used in the next levels.) The choice of $\omega_{c}$s depends on the application. For instance, conventional frequency bands, Delta, Theta, Alpha, Beta, Gamma which split frequencies of $0-100$ Hz into five non-overlapping consecutive bands, are typically used in EEG signal processing contexts.%  
Each of the filtered data is fed into hierarchical feature extractor level 2.%, shown in Fig.~\ref{fig.FEL2}
 
\noindent 
\textbf{Level 1}: Fig.~\ref{fig.FEL2} provides an overview of the second level structure. Let $\mathbf{X}^{(1,c)} \in \mathbb{R}^{S\times T_{1}}$ with $T_{1}=T_{0}$ be the input to this layer. Each $\mathbf{X}^{(1,c)}$ is partitioned in time, into $K$ (non)overlapping time windows with stride $r$. Let $\mathbf{X}^{(1,c,k)} = P_k(\mathbf{X}^{(1,c)}) $ where $P_k: \mathbb{R}^{S\times T_1}\rightarrow \mathbb{R}^{S\times (T_1/K)}, \quad k \in [K]$ is:
\begin{align}
P_{k+1}(\mathbf{X}^{(1,c)} )\triangleq
 \begin{bmatrix}
{x}^{(1,c)}_{1,kr+1} & {x}^{(1,c)}_{1,kr+2} & \!\!\!\!\!\!\cdots & {x}^{(1,c)}_{1,kr+T_1/K} \\
\cdots & \cdots & \!\!\!\!\!\!\cdots& \cdots \\
{x}^{(1,c)}_{S,kr+1} & {x}^{(1,c)}_{S,kr+2} & \!\!\!\!\!\!\cdots & {x}^{(1,c)}_{S,kr+T_1/K} \\
\end{bmatrix}
\end{align}
Each $\mathbf{X}^{(1,c,k)}$, along with the whole data $\mathbf{X}^{(1,c)}$, are fed to topology feature extractors from which topology-based features (cf. Section \ref{sec.Feature_extraction}) are extracted.  To provide the coarse data for level 2, we temporally down sample each row of the input $\mathbf{X}^{(2,c)} = f_{D} (\mathbf{X}^{(1,c)}) $ using $f_{D}: \mathbb{R}^{S\times T_1} \rightarrow \mathbb{R}^{S\times T_2} $: %time-window 
\begin{align}
f_{D}(\mathbf{X}^{(1,c)} ) \triangleq 
 \begin{bmatrix}
{x}^{(1,c)}_{1,1} &\!\!\! {x}^{(1,c)}_{1,\left\lfloor\frac{T_1}{T_2}\right\rfloor+1 } &\!\! {x}^{(1,c)}_{1,2\left\lfloor \frac{T_1}{T_2}\right\rfloor +1} &\!\!\!\!\!\!\!\!\! \cdots &\!\!\! {x}^{(1,c)}_{1,T_1} \\
\cdots &\!\!\! \cdots &\!\! \cdots&\!\!\!\!\!\!\!\!\!\! \cdots &\!\!\! \cdots\\
{x}^{(1,c)}_{S,1} &\!\!\! {x}^{(1,c)}_{S,\left\lfloor \frac{T_1}{T_2}\right\rfloor+1}  &\!\!\! {x}^{(1,c)}_{S,2\left\lfloor\frac{T_1}{T_2}\right\rfloor+1} &\!\!\!\!\!\!\!\!\! \cdots &\!\!\!\! {x}^{(1,c)}_{S,T_1} \\
\end{bmatrix}
\end{align}
where $\left\lfloor y \right\rfloor$ is the greatest integer less than or equal to $y$.

\noindent 
\textbf{Level 2}: The input of level 2, $\mathbf{X}^{(2,c)}$, provides a coarse version of the raw signal over time. Such down sampled signal enables feature extraction blocks to better capture long-term dependencies even with small lag $L$. The topology feature extraction block outputs the spatiotemporal adjacency matrix \eqref{total_W}. This matrix is used to construct the total spatiotemporal graph from which GSP-based features are derived. 
\section{Feature Extraction (Embedding)} \label{sec.Feature_extraction}
In the following we define the two feature extraction building blocks, topology and GSP-based feature extractors. 
\subsection{Topology features} 
\label{sec.topology}
We perform topology extraction at all levels. The first step is to map the tensor \eqref{tensor} to a low dimensional adjacency matrix space.
Define:
\begin{align}
\mathcal{R}_{i,j}(l) \triangleq 
\left\{ 
  \begin{array}{l l}
   \frac{1}{T} \displaystyle\sum_{k \in [T] } A_k(i,j) & \quad \mathrm{if} \quad l = 0 \\
    \frac{1}{T}\displaystyle\sum_{k \in [T] }B_{k,l}(i,j)   & \quad \mathrm{if} \quad l \in [L].  \\
  \end{array} \right.
\end{align}
Comparison with \eqref{A_def} and \eqref{B_def} reveals that $\mathcal{R}_{i,j}$ is the empirical autocovariance of signal magnitudes for each spatial vertex pairs $i,j$. Defining the adjacency matrix as $\tilde{{W}} \in \mathbb{R}^{S\times S} $:
\begin{align}\label{W_tilde_kappa}
\tilde{W}_{i,j} =\left\{ 
  \begin{array}{l l}
   1 &  \quad   \mathrm{if} \quad \left[\frac{1}{L+1}\displaystyle\sum_{l \in \left\lbrace 0,\cdots , L \right\rbrace} \mathcal{R}_{i,j}(l)\right] >\kappa  \\
   0  &  \quad  \mathrm{otherwise}   \\    
  \end{array} \right.
\end{align}
yields a spatial matrix where the connectivity of two vertices is the average empirical autocovariance of signal magnitudes over all lags and a threshold $\kappa$ is used to realize a spatial unweighted graph. Similar to the thresholded Gaussian kernel \cite{main_GSP}, the thresholding throws away the least weight edges. We thereby convert the weighted spatiotemporal graph into an unweighted spatial graph to make the graph less dense and to focus in on the topology.
%The threshold $\kappa$ is an important hyperparameter that strongly affects topological features. In the application considered in this paper, we observed that the average value of the learned weights in a randomly selected ictal case is a good choice, i.e. it gives good performance when extracted features are used for classification (details in Section \ref{sec.numerical}). %Also, different thresholds must be used in feature extraction blocks of different levels. 
Based on the graph weights \eqref{W_tilde_kappa}, we calculate the following metrics related to the graph topology: network density, local efficiency, number of connected components, size of largest component, average number of neighbours, average weights, number of self loops, characteristic path length, mean eccentricity, radius, diameter. See \cite{top_measures} for the definitions of the quantities.
\subsection{GSP-based features}\label{sec.GSP_features}
In the previous section, we examined the topology of the graph. We now consider the graph signal. The adjacency matrix for computations in this part is \eqref{total_W}, using $A_k$ \eqref{A_def} and $B_{k,l}$ \eqref{B_def}. 

\noindent
\textbf{Energy in different graph frequency bands:} 
The graph Laplacian matrix $\mathcal{L}$ is widely used in literature and is defined as
$\mathcal{L} = D - W$ (see~\cite{main_GSP}), where $D = \mathrm{ diag} (W\mathbbm{1}_{|\mathcal{V}|})$ is a diagonal matrix of the sum the weights of a vertex's neighbours.
The importance of this form of the graph Laplacian in GSP originates from the fact that its quadratic form \cite{main_GSP} evaluates variation in the graph signal across neighboring vertices which can be interpreted as graph frequencies. 
%\begin{align}\label{Laplacian_variation}
%\mathbf{x^\prime} \mathcal{L} \mathbf{x} =  \sum_{u,v \in \mathcal{N}_u} W_{u,v}[x(u)-x(v)]^2
%\end{align}
The graph Fourier transform of signal $\mathbf{x}$ in graph frequency bin $\lambda_n$, is defined as $\hat{{x}}(\lambda_n) = \mathbf{x}^\prime \mathbf{u}_n$ where $\lambda_n , \mathbf{u}_n$ for $n \in [|\mathcal{V}|] $, are respectively the eigenvalues and eigenvectors of the graph Laplacian.
We split the graph frequencies into $M$ bands: $ \Lambda_m = [b_{m-1},b_{m}), \forall m \in [M]$ where $ b_{m} \in \mathbb{R}^{+}$. The energy of $\hat{\mathbf{x}}$ in the $m$th graph frequency band is:
\begin{align}
E_m = \sum_{\lambda_n \in \Lambda_m } |\hat{{x}}(\lambda_n)|^2.
\end{align}

\noindent
\textbf{Min, max and average of eigenvalues: } GSP frequencies are discrete bins which vary with the graph Laplacian. Hence, their corresponding features, such as min, max and average, can be informative. %However, in contrast to discrete-time Fourier series (DTFS) where the frequency is defined between two static points $(-\pi,\pi]$, the GSP frequency bin values

\noindent
\textbf{Spectral graph wavelet transform} \cite{main_GSP,Wavelets_GSP}: This transformation projects the signals on the spectral graph wavelet domain. We choose the projected values that correspond to a fixed number ($z$) of maximum and minimum frequencies to be features.

\noindent
\textbf{Graph Laplacian quadratic form:} It is defined as $\mathbf{x^\prime} \mathcal{L} \mathbf{x}$ and captures the variation of signal on the graph. %defined in \eqref{Laplacian_variation}, 
\section{Numerical results} \label{sec.numerical}
To test our ideas, we used the Kaggle data set for seizure detection competition \cite{kaggle_data}. We compared our results to those of the winning solution. The winner's code is publicly available \cite{hill_winning_kaggle}. 
Since this is a binary classification problem, AUC (area under the receiver operating characteristic (ROC) curve) is used as the performance metric. 

We implemented the random forest classifier \cite{breiman2001random} which yielded the best results for both the Kaggle winner's and for our approach. The data set consists of eleven subjects (seven patients and four dogs), including a total of $2297$ ictal and $21735$ interictal labeled samples, each of $1$ second length. Half of the labeled data for each subject (patient or dog) is used for training and the other half for testing. Features are extracted according to the proposed system described in previous sections. The most significant features are selected and fed into the classifier. The frequency boundaries of level 0's bandpass filters are selected to be $(0,7,10,12,18,24,30,100,5000)$ Hz. This choice follows the conventional Delta, Theta, Alpha, Beta and Gamma bands. We also split the Beta band into subbands to provide increased resolution. The last band, from $100-5000$ Hz, while being outside the aforementioned conventional bands, is also implemented. Our tests show the inclusion of the last band improves performance. However, in comparison to other bands, the contribution of this last band is small. Our tests show that selecting smaller lags ($L=1$) for levels $0,1$ and larger lags ($L=10$) for level $2$ provide better results. %The threshold 
\begin{figure}
\centering 
\includegraphics[width=8cm]{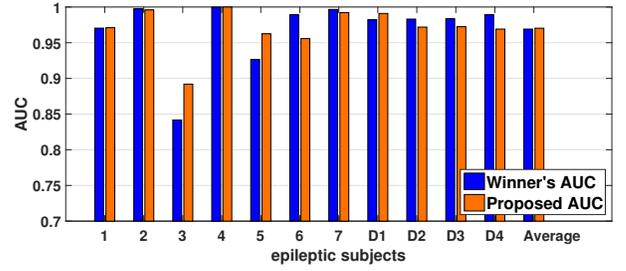}%{performance_results_zoomed.png}
\caption{Comparison of the proposed and the Kaggle winner's AUC for seven patients and four dogs. The two most improved subjects are patients 3 and 5 where the proposed approach increased AUC from $0.842$ to $0.892$ and from $0.927$ to $0.963$, respectively.}
\label{performance_results} 
\end{figure}

\begin{figure}
\centering 
\includegraphics[width=8cm]{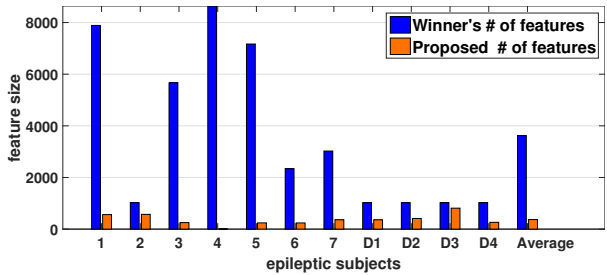}%{feature_comparison.png}
\caption{Comparison of the proposed and the Kaggle winner's number of features for seven patients and four dogs (Patient 4 has 5 features).}
\vspace{-0.2in}
\label{feature_comparison} 
\end{figure}

Fig.~\ref{performance_results} compares the classification performance across subjects. It is observed that our algorithm shows significant improvement in the two worst subjects (Patient 3 and 5). Our results for most of the other subjects are similar to the Kaggle competition winner's. Importantly, as shown in Fig.~\ref{feature_comparison}, our approach reduces the number of features significantly. The average feature size reduction is $75$ percent. Feature size is an important factor for implantable medical devices' hardware implementations, such as a closed-loop neurostimulator \cite{kassiri2017closed,kassiri2017rail, shulyzki2015320}, where memory constraints are important. Reducing the number of features, significantly reduces chip area, a desirable and sometimes essential aspect. %,abdelhalim201364 , 

\section{conclusions} \label{sec.conclusion}
We proposed a hierarchical system to extract features from spatiotemporal signals based on both the topology of the graph on which the signals are defined, and frequency-based graph signal notions. Down sampling, partitioning and filtering enabled us not only to process high dimensional input data, but also to combine the temporal, spectral and spatial characteristics of the signal. By implementing classification of an epileptic seizure detection data set, we exemplified the efficacy of the proposed framework. In comparison to the Kaggle competition winner's solution, a great reduction in the number of features for classification was achieved.
\bibliographystyle{IEEEtran}
\bibliography{references.bib}
\end{document}

%% file: styles/gww_defs.tex
\usepackage[tbtags]{amsmath} % defines many math commands and
\usepackage{amssymb}  % get, among others, blackboard bold fonts
\usepackage{verbatim} % get comment environment, + new verbatim 
\newcounter{actr}
{\begin{list}{(\alph{actr})}{\usecounter{actr}}}{\end{list}}

\newcounter{ictr}
{\begin{list}{(\roman{ictr})}{\usecounter{ictr}}}{\end{list}}

\newcommand{\qed}{\rule[0.1ex]{1.4ex}{1.6ex}}

%% file: styles/defs.tex
\newcounter{psctr}
\newcounter{probctr}[psctr]

\DeclareMathAlphabet{\mathbsf}{OT1}{cmss}{bx}{n}% bold sans serif
\DeclareMathAlphabet{\mathssf}{OT1}{cmss}{m}{sl}% slanted sans serif

\DeclareSymbolFont{bsfletters}{OT1}{cmss}{bx}{n}  
\DeclareSymbolFont{ssfletters}{OT1}{cmss}{m}{n}
\DeclareMathSymbol{\bsfGamma}{0}{bsfletters}{'000}
\DeclareMathSymbol{\ssfGamma}{0}{ssfletters}{'000}
\DeclareMathSymbol{\bsfDelta}{0}{bsfletters}{'001}
\DeclareMathSymbol{\ssfDelta}{0}{ssfletters}{'001}
\DeclareMathSymbol{\bsfTheta}{0}{bsfletters}{'002}
\DeclareMathSymbol{\ssfTheta}{0}{ssfletters}{'002}
\DeclareMathSymbol{\bsfLambda}{0}{bsfletters}{'003}
\DeclareMathSymbol{\ssfLambda}{0}{ssfletters}{'003}
\DeclareMathSymbol{\bsfXi}{0}{bsfletters}{'004}
\DeclareMathSymbol{\ssfXi}{0}{ssfletters}{'004}
\DeclareMathSymbol{\bsfPi}{0}{bsfletters}{'005}
\DeclareMathSymbol{\ssfPi}{0}{ssfletters}{'005}
\DeclareMathSymbol{\bsfSigma}{0}{bsfletters}{'006}
\DeclareMathSymbol{\ssfSigma}{0}{ssfletters}{'006}
\DeclareMathSymbol{\bsfUpsilon}{0}{bsfletters}{'007}
\DeclareMathSymbol{\ssfUpsilon}{0}{ssfletters}{'007}
\DeclareMathSymbol{\bsfPhi}{0}{bsfletters}{'010}
\DeclareMathSymbol{\ssfPhi}{0}{ssfletters}{'010}
\DeclareMathSymbol{\bsfPsi}{0}{bsfletters}{'011}
\DeclareMathSymbol{\ssfPsi}{0}{ssfletters}{'011}
\DeclareMathSymbol{\bsfOmega}{0}{bsfletters}{'012}
\DeclareMathSymbol{\ssfOmega}{0}{ssfletters}{'012}

%% file: main.bbl
% Generated by IEEEtran.bst, version: 1.13 (2008/09/30)
\begin{thebibliography}{10}
\providecommand{\url}[1]{#1}
\csname url@samestyle\endcsname
\providecommand{\newblock}{\relax}
\providecommand{\bibinfo}[2]{#2}
\providecommand{\BIBentrySTDinterwordspacing}{\spaceskip=0pt\relax}
\providecommand{\BIBentryALTinterwordstretchfactor}{4}
\providecommand{\BIBentryALTinterwordspacing}{\spaceskip=\fontdimen2\font plus
\BIBentryALTinterwordstretchfactor\fontdimen3\font minus
  \fontdimen4\font\relax}
\providecommand{\BIBforeignlanguage}[2]{{%
\expandafter\ifx\csname l@#1\endcsname\relax
\typeout{** WARNING: IEEEtran.bst: No hyphenation pattern has been}%
\typeout{** loaded for the language `#1'. Using the pattern for}%
\typeout{** the default language instead.}%
\else
\language=\csname l@#1\endcsname
\fi
#2}}
\providecommand{\BIBdecl}{\relax}
\BIBdecl

\bibitem{main_GSP}
D.~Shuman, S.~Narang, P.~Frossard, A.~Ortega, and P.~Vandergheynst, ``The
  emerging field of signal processing on graphs: Extending high-dimensional
  data analysis to networks and other irregular domains,'' \emph{IEEE Signal
  Proc. Magazine}, vol.~30, no.~3, pp. 83--98, 2013.

\bibitem{new_survey}
A.~Ortega, P.~Frossard, J.~Kova{\v{c}}evi{\'c}, J.~Moura, and P.~Vandergheynst,
  ``Graph signal processing,'' \emph{arXiv preprint arXiv:1712.00468}, 2017.

\bibitem{kassiri2017closed}
H.~Kassiri, S.~Tonekaboni, M.~T. Salam, N.~Soltani, K.~Abdelhalim, J.~L.~P.
  Velazquez, and R.~Genov, ``Closed-loop neurostimulators: A survey and a
  seizure-predicting design example for intractable epilepsy treatment,''
  \emph{IEEE trans. on biomedical circuits and systems}, vol.~11, no.~5, pp.
  1026--1040, 2017.

\bibitem{spectral_power_features}
M.~Bandarabadi, C.~Teixeira, J.~Rasekhi, and A.~Dourado, ``Epileptic seizure
  prediction using relative spectral power features,'' \emph{Clinical
  Neurophysiology}, vol. 126, no.~2, pp. 237--248, 2015.

\bibitem{hill_winning_kaggle}
\BIBentryALTinterwordspacing
M.~Hills. (2014) Winning code of {K}aggle seizure detection contest. [Online].
  Available: \url{https://github.com/MichaelHills/seizure-detection.}
\BIBentrySTDinterwordspacing

\bibitem{spatial_epilepsy}
M.~Le~Van~Quyen, J.~Martinerie, V.~Navarro, M.~Baulac, and F.~Varela,
  ``Characterizing neurodynamic changes before seizures,'' \emph{Journal of
  Clinical Neurophysiology}, vol.~18, no.~3, pp. 191--208, 2001.

\bibitem{concat_epilepsy}
A.~Shoeb and J.~Guttag, ``Application of machine learning to epileptic seizure
  detection,'' in \emph{Proc. of the 27th Int. Conf. on Machine Learning},
  2010, pp. 975--982.

\bibitem{PLV}
N.~Caramia, F.~Lotte, and S.~Ramat, ``Optimizing spatial filter pairs for {EEG}
  classification based on phase-synchronization,'' in \emph{IEEE Int. Conf.
  Acoustics, Speech, Signal Proc.}, 2014, pp. 2049--2053.

\bibitem{dynamic_graphs}
D.~Romero, V.~Ioannidis, and G.~Giannakis, ``Kernel-based reconstruction of
  space-time functions on dynamic graphs,'' \emph{IEEE Selected Topics in
  Signal Proc.}, vol.~11, no.~6, pp. 856--869, 2017.

\bibitem{DGFT}
M.~Villafa{\~n}e-Delgado and S.~Aviyente, ``Dynamic graph fourier transform on
  temporal functional connectivity networks,'' in \emph{IEEE Int. Conf.
  Acoustics, Speech, Signal Proc.}, 2017, pp. 949--953.

\bibitem{SVARM_brain_connectivity}
Y.~Shen, B.~Baingana, and G.~B. Giannakis, ``Topology inference of directed
  graphs using nonlinear structural vector autoregressive models,'' in
  \emph{IEEE Int. Conf. Acoustics, Speech, Signal Proc.}, 2017, pp. 6513--6517.

\bibitem{journal_graph_laplacian_learning}
H.~Egilmez, E.~Pavez, and A.~Ortega, ``Graph learning from data under laplacian
  and structural constraints,'' \emph{IEEE Selected Topics in Signal Proc.},
  vol.~11, no.~6, pp. 825--841, 2017.

\bibitem{functional_connectivity}
B.~Biswal, F.~Zerrin~Yetkin, V.~Haughton, and J.~Hyde, ``Functional
  connectivity in the motor cortex of resting human brain using echo-planar
  {MRI},'' \emph{Magnetic Resonance in Medicine}, vol.~34, no.~4, pp. 537--541,
  1995.

\bibitem{diffusion_graph_learning}
B.~Baingana and G.~Giannakis, ``Tracking switched dynamic network topologies
  from information cascades,'' \emph{IEEE Trans. on Signal Proc.}, vol.~65,
  no.~4, pp. 985--997, 2017.

\bibitem{top_measures}
M.~Rubinov and O.~Sporns, ``Complex network measures of brain connectivity:
  uses and interpretations,'' \emph{Neuroimage}, vol.~52, no.~3, pp.
  1059--1069, 2010.

\bibitem{Wavelets_GSP}
D.~Hammond, P.~Vandergheynst, and R.~Gribonval, ``Wavelets on graphs via
  spectral graph theory,'' \emph{Applied and Comput. Harmonic Analysis},
  vol.~30, no.~2, pp. 129--150, 2011.

\bibitem{kaggle_data}
\BIBentryALTinterwordspacing
(2014) {K}aggle seizure detection data set. [Online]. Available:
  \url{https://www.kaggle.com/c/seizure-detection/data.}
\BIBentrySTDinterwordspacing

\bibitem{breiman2001random}
L.~Breiman, ``Random forests,'' \emph{Machine learning}, vol.~45, no.~1, pp.
  5--32, 2001.

\bibitem{kassiri2017rail}
H.~Kassiri, M.~T. Salam, M.~R. Pazhouhandeh, N.~Soltani, J.~L.~P. Velazquez,
  P.~Carlen, and R.~Genov, ``Rail-to-rail-input dual-radio 64-channel
  closed-loop neurostimulator,'' \emph{IEEE Journal of Solid-State Circuits},
  vol.~52, no.~11, pp. 2793--2810, 2017.

\bibitem{shulyzki2015320}
R.~Shulyzki, K.~Abdelhalim, A.~Bagheri, M.~T. Salam, C.~M. Florez, J.~L.~P.
  Velazquez, P.~L. Carlen, and R.~Genov, ``320-channel active probe for
  high-resolution neuromonitoring and responsive neurostimulation,'' \emph{IEEE
  trans. on biomedical circuits and systems}, vol.~9, no.~1, pp. 34--49, 2015.

\end{thebibliography}
